\documentclass[twocolumn,showpacs,prl]{revtex4}

\newcommand{\BE}{\begin{equation}}
\newcommand{\EE}{\end{equation}}
\newcommand{\BA}{\begin{eqnarray}}
\newcommand{\EA}{\end{eqnarray}}

\begin{document}

\title{Localization of classical waves in two dimensional random
media: A comparison between the analytic theory and exact
numerical simulation}
\author{Bikash C. Gupta and Zhen Ye}
\affiliation{Wave Phenomena Laboratory, Department of Physics,
National Central University, Chungli, Taiwan 32054}
\date{\today}


\begin{abstract}

The localization length for classical waves in two dimensional
random media is calculated exactly, and is compared with the
theoretical prediction from the previous analytic theory.
Significant discrepancies are observed. It is also shown that as
the frequency varies, critical changes in the localization
behavior can occur. Possible reasons for the discrepancies are
discussed.

\end{abstract}


\maketitle


\section{Introduction}

Propagation of waves through random medium has been and continues
to be a subject of vivid research \cite{fol,lax,wat,twe,fri,pag}.
When propagating through media containing many scatterers, waves
will be repeatedly scattered by each scatterer, forming a multiple
scattering process\cite{ish}. Multiple scattering is responsible
for phenomena such as random laser \cite{law,Zhang}, electronic
transport in impure solids \cite{and}, and photonic or acoustic
bandgaps \cite{yab,san,kus}. Under proper conditions, multiple
scattering leads to the unusual phenomenon of wave localization, a
concept introduced by Anderson \cite{and}.

Such a localization phenomenon has been characterized by two
levels. One is the weak localization associated with the enhanced
back scattering due to constructive interference from the reversed
propagating paths. The second is termed as the strong localization
(for brevity just termed as localization hereafter), in which
significant inhibition of transmission surfaces, indicating that
the energy mostly remains in a region of space in the neighborhood
of the emission. The generally accepted wisdom on the connection
between the weak localization and the strong localization is that
enhanced backscattering of a diffusive wave packet leads to an
effective reduction in the diffusion constant of the wave packet.
When the influence of the increased backscattering is so
overwhelming that the diffusion constant vanishes, the strong
localization sets in.

It is worthy to mention that tremendous efforts have been devoted
to investigate the localization phenomenon for classical waves in
random media over the past several years
\cite{kir,gen,dal,lag,van,McCall2,aad,cjp,Nat,Sridhar}. However,
observation of classical wave localization is a difficult task,
partially because suitable systems are hard to find and partially
because observation is often complicated by such effects as
absorption and attenuation.

One important quantity associated with the wave localization is
the localization length which is defined as the characteristic
decay length of the transmitted intensity through the system from
the source. Taking into account of the enhanced backscattering,
the corrected (renormalized) diffusion constant may be obtained
which then may be set to zero to obtain the localization length.
This theoretical version of the localization length may be found
in the literature for various dimensions for both the electronic
waves and classical waves (e.~g.\cite{kir,DV}, and summarized in
Ref.~\cite{pin}).

A question we raise in this paper is about the validity of the
existing theoretical formula for localization length, and
subsequently the appropriateness or accuracy of the previous
theory of wave localization. A main reason behind raising this
question is that it would be helpful to experimentalists in
designing experiments, provided that the theoretical formula for
localization length is accurate enough in both qualitative and
quantitative sense, and otherwise, it may mislead the
experimentalists if the theory is not accurate enough.

The main aim of this paper is to answer the questions raised above
for classical waves. One possible way is to concentrate on an
exactly computable system and find the exact characteristic decay
length of the transmitted intensity with the sample size, and
compare with the theoretical value of this particular system
obtained from the existing theory. In this paper, we first outline
the previous theory and show a numerically exact procedure for the
calculation of the localization length for classical waves in two
dimensional media. Then we compare the two results, to gain
information about the accuracy and appropriateness of the previous
theory. In addition, we will also investigate the localization
behavior as the frequency varies. These will be followed by a
detailed discussion on the implications of the results.

\section{Formalism}

\subsection{Existing theory}

Here we briefly review the existing theory for wave localization.
As wave propagates in random media, it experiences multiple random
scattering, and as a result, the wave loses its propagating phase,
leading to the gradual decreases of the coherence of the wave in
the absence of absorption, i.~e. elastic scattering. Meanwhile,
diffusive wave is built up as more and more scattering takes
place. The traditional picture of localization is in fact a
version of localization of diffusive waves. In other words, the
conventional wisdom towards the localization mechanism is the
absence of diffusion. The procedure to obtained the localization
state can be briefly summarized as follows.

The quantity, $D^{(B)}$ which is a measure of diffusion of
classical waves is called the classical Boltzman diffusion
constant and it may be derived under the coherent potential
approximation \cite{lax}, and is given as \cite{pin}
\begin{equation}
D^{(B)} \sim \frac{v_t l}{d_{dim}}
\end{equation}
where $v_t$ is the transport velocity, $l$ is the mean free path
and $d_{dim}$ is the dimensionality.

As waves scattered along any two reversed paths in the backward
direction interfere constructively, leading to the enhanced
backscattering effect, which will then add corrections to the
diffusion coefficient. In the two dimension case, such an enhanced
backscattering effect is represented by a set of maximally crossed
ladder diagrams. An evaluation of these diagrams leads to an
integration for which two cut-off limits have to be introduced to
avoid the divergence. In this way, the correction to the diffusion
constant for two dimensional system may be found to be \cite{pin}
\begin{equation}
\delta D \sim  - {\rm ln}(L_M/l_m) \label{eq:correction}
\end{equation}
where $L_M$ and $l_m$ are the two cut-off limits. It is then
interpreted in the previous theory that the cut-off limit $l_m$ is
a measure of the minimum scaling for the waves and is thought to
be related to (for example) the mean free path, whereas $L_M$ is a
measure of the effective size of the sample. It is rather
important to note that the correction in Eq.~(\ref{eq:correction})
is not only negative but diverges as $L_M$ approach infinity. This
is obviously unphysical, since the conductance or the corrected
diffusion constant cannot be negative. To avoid the problem, it
was conjectured that $L_M$ is in fact related to the localization
range, or simply the localization length, in such way that when
$L_M$ is equal to the localization length denoted by $\xi$ say
(i.~e. $L_M=\xi$), the corrected diffusion coefficient becomes
zero, i.~e. the absence of diffusion:
\BE D_R(\xi) = D^{(B)} +
\delta D(\xi) = 0. \label{eq:d}
\EE
The localization length $\xi$
is solved from this equation. It is obvious that this equation
always renders a solution and thus a localization length can
always be found in two dimensions. Such an absence-of-diffusion
mechanism is the core of the previous theory of localization in
two dimensions, and supports the assertion from a scaling analysis
that all waves are localized in two dimensions\cite{gang4}.

As will be shown later, there are significant discrepancies
between the result from the absence-of-diffusion mechanism and the
result from the exact numerical computation. These discrepancies
would indicate either that the theory based on the diffusion
mechanism is not accurate enough or simply that the theory has not
yet fully captured the essence of the localization. We will come
back to this later. Since the existing theory heavily depends on
the diffusion picture, we may call it the {\it diffusion based
theory}.

Equation (\ref{eq:d}) leads to the following solution for the
localization length in two dimensions
\begin{equation} \xi_{\rm theory}=l {\rm exp}(\frac{\pi}{2} {\rm
Re}[k_{\rm{eff}}] l), \label{eq:theory}
\end{equation}
Under the effective medium theory, the mean free path, $l$ may be
obtained from the characteristic decay length of the coherent
intensity and may be expressed as
$l=\frac{1}{2 {\rm Im}[k_{{\rm eff}}]}$ and the effective wave number,
$k_{\rm{eff}}$ may be written as
\begin{equation}
k_{\rm eff} = k + \sqrt{\frac{2 \pi}{i k}} \rho f(0)
\end{equation}
where $k$ is the wave vector of the incident wave propagating
through the random scattering media with the scatterer numerical
density $\rho$; $f(0)$ represents the scattering function of a
single scatterer in the forward direction. The forward scattering
function can be calculated by the standard series expansion method
for a cylinder with radius $a$ when a wave with wave vector $k$ is
incident normally on it. The forward scattering function is given
as
\begin{equation}
f(0) = \sqrt{\frac{2}{\pi k}}\sum_{n=0}^\infty C_n e^{-i(n \pi/2 +
\pi/4)}
        \end{equation}
 with
$$           C_n = \frac{-\epsilon_n (J_n^\prime(ka)
J_n(k^\prime a)
           - \frac{1}{gh} J_n(ka) J_n^\prime (k^\prime a))}
           {{H_n^{(1)}}^\prime (ka) J_n (k^\prime a)
           - \frac{1}{gh} H_n^{(1)} (ka) J_n^\prime (k^\prime a)}
$$ where $\epsilon_0 = 1$ and $\epsilon_{n \ge 1} =
2$, $g$ is the ratio of the density of the scatterer to that of
the media, $h$ is the ratio of the wave speed in the sactterer to
that in the media, $J_n$ represents the Bessel function of order
$n$, $H_n^{(1)}$ represents $n$-th order Hankel function of first
kind and prime indicates the derivative.

\subsection{Exact numerical calculation}

Thus, on one hand the localization length for classical wave in a
system consisting of any kind of cylindrical scatterers placed
randomly in any media may be evaluated by the use of equations
from the previous theory. On the other hand, the localization
length for classical waves in two dimensional random media may
also be calculated exactly by the use of the multiple scattering
technique.

Consider $N$ straight cylinders located at $\vec{r_i}$ with $i =
1, 2,...,N$ to form a {\it completely} random array. An acoustic
line source transmitting monochromatic waves is placed at $\vec
r_s$. The scattered wave from each cylinder is a response to the
total incident wave composed of the direct wave from the source
and the multiply scattered waves from other cylinders. The final
wave reaching a receiver located at $\vec r_r$ is the sum of
direct wave from the source and the scattered waves from all the
cylinders. Such scattering problem can be solved exactly,
following Twersky. While the details are in \cite{proc}, the
essential procedures are summarized below. The formulation
presented below has been successfully applied to explain the
recent experimental observations of acoustic
crystals\cite{che,che1}.

The scattered wave from the $j$-th cylinder can be written as
\begin{equation}
p_s(\vec{r}, \vec{r_j}) = \sum_{n=-\infty}^\infty i \pi A_n^j
H_n^{(1)}(k|\vec{r} - \vec{r_j}|) e^{in\phi_{\vec{r} -
\vec{r_j}}},
\end{equation}
where $k$ is the wavenumber in the medium, $H_n^{(1)}$ is the
$n$-th order Hankel function of first kind, and $\phi_{\vec{r} -
\vec{r_j}}$ is the azimuthal angle of the vector $\vec{r} -
\vec{r_j}$ relative to the positive $x$ axis. The total incident
wave around the $i$-th cylinder $(i=1,2,...,N; i \ne j)$ is
summation of the direct incident wave from the source and the
scattered waves from all other scatterers, can be expressed as
\begin{equation}
p_{in}^i(\vec r) = \sum_{n=-\infty}^\infty B_n^i J_n(k|\vec r -
\vec{r_i}|) e^{in\phi_{\vec r - \vec{r_i}}}.
\end{equation}

The coefficients $A_n^i$ and $B_n^i$ can be solved by expressing
the scattered wave $p_s(\vec r, \vec{r_j})$, for each $j \ne i$,
in terms of the modes with respect to the $i$-th scatterer by the
addition theorem for Bessel function \cite{gra}. Then the usual
boundary conditions are matched at each individual scattering
cylinder. This leads to
\begin{equation}
B_n^i = S_n^i + \sum_{j=1, j \ne i}^N C_n^{j,i}, \label{eq:6}
\end{equation}
\noindent with
\begin{equation}
S_n^i = i \pi H_{-n}^{(1)}(k|\vec{r_i}|)e^{-i n \phi_{\vec r_i}},
\end{equation}
and
\begin{equation}
C_n^{j,i} = \sum_{l=-\infty}^\infty i \pi A_l^j
          H_{l-n}^{(1)}(k|\vec{r_i} - \vec{r_j}|)
          e^{i(l-n)\phi_{\vec{r_i} - \vec{r_j}}},
\end{equation}
and
\begin{equation}
B_n^i = i \pi \tau_n^i A_n^i,
\end{equation}
where $\tau_n^i$ are the transfer matrices relating the acoustic
properties of the scatterers and the surrounding medium, and have
been given by Eq. (21) in Ref. \cite{che}.

The coefficients $A_n^i$ and $B_n^j$ can then be inverted from
Eq.~(\ref{eq:6}). Once the coefficients $A_n^i$ are determined,
the transmitted wave at any special point is given by
\begin{equation}
p(\vec r) = p_0(\vec r) + \sum_{i=1}^N \sum_{n=-\infty}^\infty
            i \pi A_n^i H_n^{(1)}(k|\vec r - \vec{r_i}|)
            e^{in\phi_{\vec r - \vec{r_i}}},
\end{equation}
where $p_0$ is the field when no scatterers are present. The
normalized transmission is defined as $T = p/p_0$ and therefore,
the acoustic intensity is represented as $|T|^2$.

The averaged localization length is subsequently determined
by\cite{Kramer} \BE {\xi}_{\rm exact} = -\frac{L}{<\ln|T(L)|^2>},
\EE where $L$ is the sample size, and $<\cdot>$ denotes the
ensemble average over the random distribution of the scatterers.
Thus obtained localization length can be compared to that in
Eq.~(\ref{eq:theory}) obtained analytically from the previous
theory.

\section{Results comparison and discussion}

\subsection{Results comparison}

The system we consider here consists of $N$ identical air
cylinders placed randomly in water medium. The reason behind
considering the system as the air cylinders in water is that due
to the large contrast of densities for air and water, and also a
large contrast of sound speeds in air and water, the air cylinders
act as strong scatters to the waves propagating in water media.
The radius of each air cylinder is $a$. The fraction of area
occupied by the cylinders per area is $\beta$. the average
distance between nearest neighbors is therefore
$d=(\frac{\pi}{\beta})^{1/2} a$, which is also the lattice
constant for the corresponding regular lattice array. All the
cylinders are placed completely randomly within a circle of radius
$L$. A transmitting line source is located at the center and the
receiver is located outside the scattering cloud. In the
computation, the acoustic intensity is normalized in such a way
that its value equals unity when there are no scatterers present.
Thus the uninteresting geometrical spreading effect is naturally
eliminated. All the lengths are scaled by the parameter $d$, and
the frequency is presented in terms of $ka$ to make the
computation non-dimensional.

Fig.~(1) shows the variation of $<\ln|T|^2>$ as a function of the
system size, $L$ for three different values of frequency ($ka$ =
0.01, 0.013 and 0.016)  with the filling fraction, $\beta=0.001$.
The $<\cdot>$ implies the configuration average of the total
transmitted intensity. Number of configurations considered here is
200. It is apparent from the Fig.~(1) that the averaged total
transmitted intensity decays exponentially with the dimensionless
system size, $L$ for all the $ka$ values shown in the figure which
in turn indicates the localization of wave at those frequencies.
The localization length is nothing but the characteristic decay
length and thus it may be obtained from the inverse of the slope
of the straight lines (in Fig.~(1)) for those frequencies.
Similarly one may obtain the localization length of classical
waves at other frequencies from the exact numerical computation.

Fig.~(2) presents the dimensionless localization length ($\xi$) as
a function of frequencies ($ka$) for $\beta=0.001$. The dashed
curve with circles represent the localization length obtained from
the exact numerical computation using the multiple scattering
technique while the solid curve represents the localization length
calculated from the theoretical formula as prescribed earlier in
Eq.~(\ref{eq:theory}). From the dashed curve we observe that the
localization length increases very rapidly as the frequency of the
wave is decreased at the low frequency regime, the localization
length also increases rapidly with the increase of frequency at
the high frequency regime and there is a minimum of localization
length at an intermediate frequency. For lower values of
frequencies the waves propagating through the random media feel
the media as homogeneous over a large distance depending on the
wavelength, for higher values of frequencies the waves are like
very localized objects that can easily propagate between the
scatterers and at an optimal frequency, the waves feel the
strongest scattering in the random media. Thus, the larger values
of localization lengths at lower and higher frequencies and the
minimum value of localization length at an intermediate frequency
are understood physically.

However, we observe that critical change in the localization
length takes place at lower and higher frequencies. On the other
hand, the solid curve also shows similar behavior. But, one
important difference is that the minimum of the dashed curve
occurs at a higher frequency when compared with that of the solid
curve. In other wards, the strongest localization predicted from
the previous theoretical formula occurs at a frequency which is
lower than that obtained from the exact numerical computation. And
as a result, we observe that the two curves cross each other at
some frequency, say $(ka)_{cr}$. For $ka < (ka)_{cr}$ the
localization length obtained from exact numerical calculation is
larger than that obtained from the existing theoretical formula
while for $ka > (ka)_{cr}$, the theoretical value dominates over
the exact value of localization length. [Note that both the
results are theoretical though, the results (values) obtained from
the previous theoretical formula are termed as theoretical results
(values) and those obtained from exact numerical calculations are
termed as exact results (values) and we may use this convention
for our convenience.] In the low and high frequency regimes, the
discrepancies between the theoretical and the exact values are
significant.

We have also done similar calculations for $\beta=0.01$ and
compared the localization length obtained from the theoretical
formula with that obtained from the exact numerical computation
and is shown in Fig.~(3). While the theoretical formula seems to
capture the shape of the exact results (Fig.~(2) and Fig.~(3)),
but there are significant discrepancies with regard to whether the
waves are most localized, and where localization length starts to
increase rapidly.

Shown in Fig.~(4) is the variation of localization length ($\xi$)
as a function of the filling factor $\beta$ for $ka=0.009$ which
is less than $(ka)_{cr}$ in Fig.~(2). The dashed curve with circle
represents the exact localization length obtained from numerical
calculation, and the solid curve is obtained from theoretical
formula. The solid curve shows the monotonous decrease of the
localization length as the filing factor increases but the exact
result (dashed curve with circle) shows completely different
behavior: the localization length initially decreases and then
increases monotonically, leaving a minimum at some value of
$\beta$. Thus, the wave with frequency $ka=0.009$ becomes
localized most strongly only for a specific value of the average
density of scatterers. However, for $ka=0.009$, the exact
localization length is larger than the theoretical values for all
filling factors. Thus here we observe that at the frequency,
$ka=0.009$ the theoretical results do not even capture the trend
of the exact results. Similar behavior can be observed for a small
range of frequencies around $ka=0.009$.

The variation of localization length ($\xi$) as a function of the
filling factor $\beta$ for a moderate frequency ka=0.015 which is
greater than $(ka)_{cr}$ in Fig.~(2) is shown in Fig.~(5). The
dashed curve with circle represents the exact result and the solid
curve represents the theoretical result for localization length.
The localization length decreases with filling factor for both the
curves. However, there is a cross over between two curves at some
value of $\beta$, say $\beta_{cr}$ below which the theoretical
value is higher than the exact value and above $\beta_{cr}$ the
exact values are larger than the theoretical values for
localization length. Thus at $ka=0.015$, the exact and theoretical
curves are similar in shape, the exact and theoretical values for
localization length matches at $\beta=\beta_{cr}$, otherwise there
are significant quantitative discrepancies. All the observed
discrepancies between the theoretical and exact results are
probably due to the approximations involved in the previous
theory.

In order to study the fluctuation behavior of localization, we
define an exponent, by analogy with one-dimensional
cases\cite{Lev}, \BE \gamma \equiv - \frac{1}{L}\ln(|T|^2),\EE
which is reciprocal to the localization length. Fig.~ 6(a) shows
the variance of $\gamma$, i.~e. $\mbox{var}(\gamma) = <\gamma^2> -
<\gamma>^2$, as a function of $ka$ for $\beta=0.001$.

We interestingly observe that for a range of frequencies, the
fluctuation is rather small indicating that the wave is strongly
localized in this range of frequencies. However, at low and high
frequencies (below and above the range of frequencies where the
fluctuation is very small) the fluctuation is high. This is an
indication of some critical change in the localization behavior
and it tends to suggest that waves become very rapidly and
critically less localized at low and high frequencies. It may also
be an implication of the localization-delocalization transition,
as discussed in\cite{emi,APL}. These cannot be explained in the
context of the current diffusion based theory. We also plot the
variance of the exponent $\gamma$ versus the mean $<\gamma>$ in
Fig.~6(b). Here we do not observe the linear dependence between
the mean localization length and the variance, similar to 1D
cases\cite{Lev,Luan}

\subsection{Discussion}

All the deviations from the predictions from the existing theory
should possibly have a couple of important implications. The first
is merely that the theory is not accurate enough, since it is
necessarily based upon a perturbation scheme given the
complications involved in the problem. If this scenario is valid,
we could believe that though inaccurate, the present theory still
captures the fundamental nature of the localization phenomenon. It
would then be expectable that the discrepancies will be reduced
when more and more relevant multiple scattering processes are
included in the theoretical derivation. A next task would thus be
to look for these scattering processes.

The second implication would be simply that the existing theory,
which is based upon the absence-of-diffusion mechanism, has not
yet fully encapsulated the essence of the phenomenon of wave
localization. Beside the results shown above, there are other
evidences indicating that this line of reasoning should not be
discarded too lightly. For example, it has been shown that in the
localized state, a genuine phase-coherence behavior prevails as a
unique feature of localized waves\cite{emi}. This coherent
behavior can be understood as follows. For quantum mechanic or
acoustic waves, the current can be written as $ \vec{J} \sim
\mbox{Re}[\psi(-i)\nabla\psi],$ where $\psi$ stands for the wave
function for quantum mechanical systems and for the pressure in
acoustic systems. Writing the field as $ \psi = |\psi|
e^{i\theta},$ the current becomes $\vec{J} \sim
|\psi|^2\nabla\theta.$ It is clear that when $\theta$ is constant
at least by domains while $|\psi| \neq 0$, the flow stops, i.~e.
$\vec{J}=0$, and the wave or the energy is localized in space,
i.~e. $|\psi|^2 \neq 0$. Obviously the constant phase $\theta$
indicates the appearance of a long range ordering in the system.
All these have been demonstrated successfully not only for two
dimensional media\cite{emi,APL}, but for one and three dimensions
as well\cite{Luan,cjp,Hsu}. It can be easily shown that the same
consideration also holds for electromagnetic waves\cite{EM}. In
the current diffusion based theory, the phase information is not
attained. Therefore, at least the theory cannot account for the
phase-coherence related localization phenomenon. Additionally, the
theory could also be expected to fail when waves come to a
complete stop before the diffusion becomes dominant.

We would like to make a further note on the existing theory of
wave localization. It has been widely believed in the literature
that the theory has been tested experimentally to be very
successful. Upon a close scrutiny, we find that the success is
mainly based upon two types of experiments. One is the measurement
of the enhanced backscattering (e.~g. Ref.~\cite{van}). In a
rigorous numerical simulation, it has been shown that the enhanced
backscattering is not directly related to the
localization\cite{aad}. Waves are not necessarily localized when a
strong enhanced backscattering exits, and sometimes waves can be
localized even when the enhanced backscattering is weak. Asides
from few exceptions\cite{McCall2,Sridhar}, the other type of
experiments is based on observations of the exponential decay of
waves as they propagate {\it through} disordered media, as
summarized in \cite{Nat}; this setup scenario is in fact also the
base for the previous scaling analysis\cite{gang4}. The recent
research\cite{Chen}, however, has indicated that while this type
of experiments can indeed help obtain information about whether a
disordered medium will or not prevent waves from propagation {\it
through} the medium, but this type of experiments is {\it not}
sufficient to discern whether the medium really only has localized
states; this discovery is consistent with the prediction of a
recent scaling analysis\cite{zhen}. Unwanted effects of
non-localization origin can also contribute to the wanted
exponential decay, likely making data interpretation ambiguous. In
other words, there is a crucial need to differentiate the
situation that waves are blocked from transmission from the
situation that waves can be actually localized in the medium. This
is also essential to a scaling analysis\cite{zhen}. It is worth
noting that there was report of the observation of microwave
localization in two dimensions when a transmitting source is
inside disordered media\cite{McCall2}. However, the diffusion
based theory has not been verified against this experimental
result. Instead, an exact numerical simulation based on the phase
coherence picture is shown to agree well with the experimental
observation\cite{EM}.

\section{Summary}

The localization length for classical waves in two dimensional
random media is obtained from exact numerical computation as well
as from the previous theoretical formula and then the results are
compared. Significant discrepancies between the two results is
observed. Furthermore, results indicate some critical change in
the localization behavior at low and high frequencies, and it
seems to suggest that waves become very rapidly and critically
less localized at these frequencies. These features are not
readily seen in the previous theoretical picture of wave
localization. Therefore it is suggested that caution should be
taken when applying the previous analytic theory to infer the
localization length of classical waves. Possible implications of
the present research are discussed.

\section*{Acknowledgment}

This work received support from National Science Council of
Republic of China (Grant No: NSC 90-2811-M008-004).

\section*{Figure Captions}

\begin{description}

\item[Fig. 1]
$<\ln|T|^2>$ is plotted as a function of sample size ($L$) for
$ka=0.010$ (circles), $ka=0.013$ (stars) and $ka=0.016$ (diamonds)
respectively. Here the filling factor ($\beta$) is 0.001.

\item[Fig. 2] Localization length
($\xi$) is shown as a function of frequency ($ka$) for
$\beta$=0.001. The dashed curve with circles represent the exact
values obtained numerically while the solid curve is obtained from
theoretical formula.

\item[Fig. 3] Same as the Fig.~2
except that $\beta$ =0.01 here.

\item[Fig. 4] Localization length
($\xi$) is shown as a function of filling factor ($\beta$) for
$ka$=0.009. The dashed curve with circles represent the exact
values obtained numerically while the solid curve is obtained from
theoretical formula.

\item[Fig. 5] Same as the Fig.~4
except that $ka=0.015$ here.

\item[Fig. 6] (a) This shows the variance of the localization exponent
$\mbox{var}(\gamma)$ as a function of frequency ($ka$) with
$\beta=0.001$. (b) The variance $\mbox{var}(\gamma)$ is shown as a
function of the mean $<\gamma>$ for $\beta=0.001$.

\end{description}

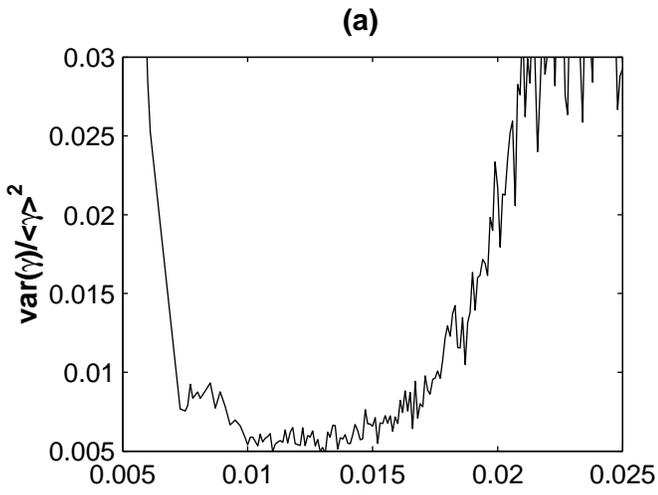
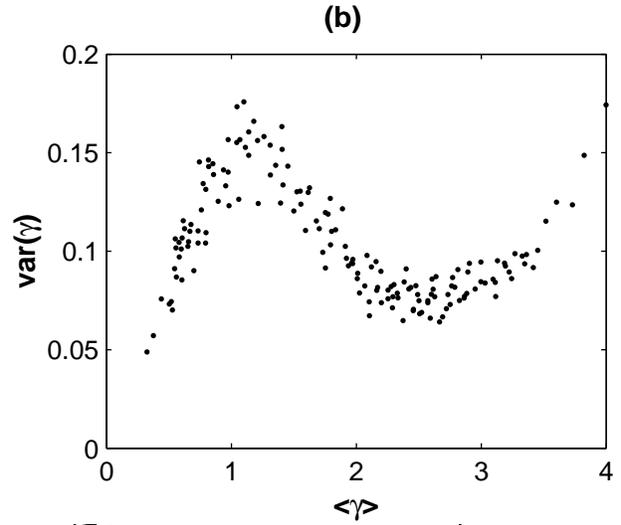
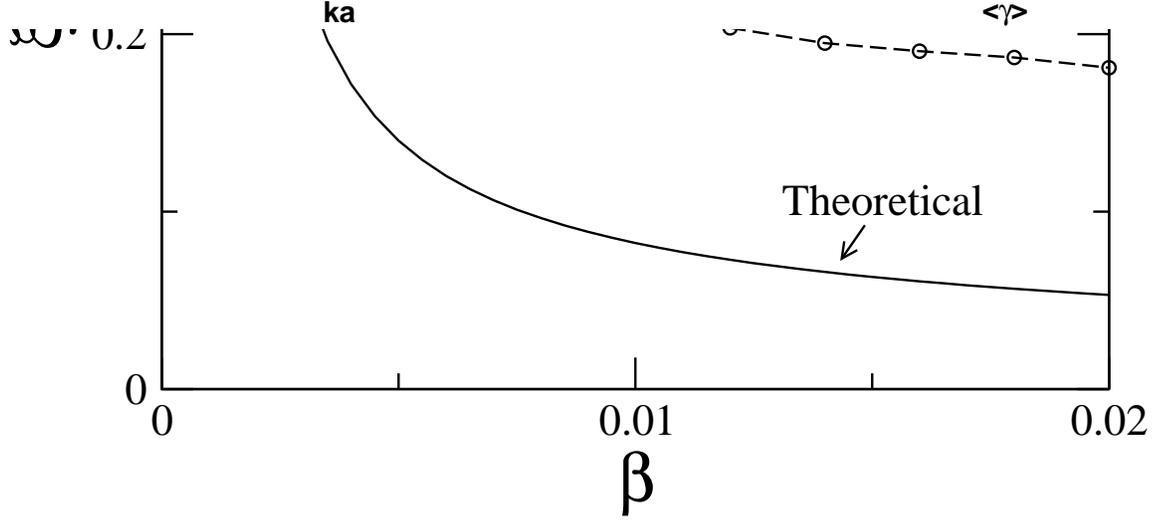
\end{document}